\documentclass[]{easychair}

\usepackage{doc}
\usepackage{makeidx}
\usepackage{pst-node,float,multicol}
\usepackage{fancyvrb,keyval,ifthen,hhline,stmaryrd}
\usepackage[vlined,ruled]{algorithm2e}
\usepackage{amssymb}
\usepackage{amsmath,alltt}

\newcommand{\tensor}{\otimes}
\newcommand{\dsum}[2]{\bigoplus_{#1}{#2}}
\newcommand{\ssum}[2]{\sum_{#1}{#2}}
\newcommand{\id}[1]{I_{#1}}

\newcommand{\shift}[2]{J_{#1}^{#2}}
\newcommand{\hdp}{\mid}
\newcommand{\Complex}{\mathbb{C}}
\newcommand{\Real}{\mathbb{R}}
\newcommand{\Dirac}{Dirac}

\newcommand{\isInvertible}[1]{\text{invertible}(#1)}

\newcommand{\Print}[1]{\text{print}}

\newcommand{\U}{U}

\newenvironment{template}[1]{%
\renewcommand{\algorithmcfname}{Algo}
\begin{algorithm}[H]
\label{#1}
\SetKwInOut{Constant}{Constant}
\SetKwInOut{Input}{Input}
\SetKwInOut{Output}{Output}
\SetKwInOut{Match}{Match}
\SetKwInOut{Require}{Require}
\SetKwInOut{Var}{Var}
}{%
\end{algorithm}
}

\newcommand*{\dquote}[1]{``#1''}

\begin{document}

%
\title{QIRAL: A High Level Language for Lattice QCD Code Generation}

%
\titlerunning{QIRAL: A High-Level Language for LQCD Code Generation}




%
\author{
    Denis Barthou\\
    \affiliation{University of Bordeaux,}\\
    \affiliation{Bordeaux, France}\\
    \affiliation{\url{denis.barthou@labri.fr}}\\
\and
    Gilbert Grosdidier\\
    \affiliation{Laboratoire de l'Acc\'elerateur Lin\'eaire,}\\
\affiliation{Orsay, France}\\
\affiliation{\url{grosdid@in2p3.fr}}\\
\and
    Michael Kruse\\
    \affiliation{INRIA Saclay / ALCHEMY,}\\
    \affiliation{Orsay, France} \\
    \affiliation{\url{michael.kruse@inria.fr}}\\
\and
    Olivier Pene\\
    \affiliation{Laboratoire de Physique Th\'eorique,}\\
    \affiliation{Orsay, France}\\
    \affiliation{\url{olivier.pene@th.u-psud.fr}}\\
\and
    Claude Tadonki\\
    \affiliation{Mines ParisTech/CRI - Centre de Recherche en Informatique,}\\
    \affiliation{Fontainebleau, France}\\
    \affiliation{\url{claude.tadonki@mines-paristech.fr}}
}
\authorrunning{Barthou \textit{et al.}}
\indexedauthor{Barthou, Denis}

\maketitle
%
\begin{abstract}
  Quantum chromodynamics (QCD) is the theory of subnuclear
physics, aiming at modeling the strong nuclear
force, which is responsible for the interactions of nuclear
particles. Lattice QCD (LQCD) is the corresponding discrete formulation, widely used for simulations. The computational demand for the LQCD
  is tremendous. It has played a role in the history of
  supercomputers, and has also helped
   defining their future. Designing efficient 
  LQCD codes that scale well on large (probably hybrid) supercomputers requires to express many levels of parallelism,
  and then to explore different algorithmic solutions. While
  algorithmic exploration is the key for efficient parallel codes,
  the process is hampered by the necessary coding effort.

  We present in this paper a domain-specific language, QIRAL, for a
  high level expression of parallel algorithms in LQCD. Parallelism is
  expressed through the mathematical structure of the sparse matrices
  defining the problem. We show that from these expressions and from
  algorithmic and preconditioning formulations, a parallel code can be
  automatically generated. This separates algorithms and mathematical
  formulations for LQCD (that belong to the field of physics) from the
  effective orchestration of parallelism, mainly related to compilation and optimization for parallel architectures.
\end{abstract}

\section{Introduction}
Quantum Chromodynamics (QCD) is the theory of strong subnuclear
interactions\cite{qcd}.  Lattice QCD (LQCD) is the numerical approach
to solve QCD equations.  LQCD simulations are extremely demanding in
terms of computing power, and require large parallel and distributed
machines. At the heart of the simulation, there is an inversion problem:
\begin{equation}
\label{eq:dirac}
A x = b,
\end{equation}
 where $A$ is a large sparse (also sparse and implicit) matrix, called Dirac matrix,
$b$ is a known vector and $x$ is the unknown. To model reality
accurately, $A$ should be of size around $2^{32}\times 2^{32}$
elements at least. Current simulations on supercomputers handle sizes
up to $2^{24}\times 2^{24}$ elements.  High performance LQCD codes on
multicores, multinode and hybrid (using GPUs) architectures are quite
complex to design: performance results from the interplay between the
algorithms chosen to solve the inversion and the parallelism
orchestration on the target architecture. Exploring the space of
algorithms able to solve the inversion, such as mixed precision or
communication-avoiding algorithms, aggressive preconditioning,
deflation techniques or any combination of these is essential in order
to reach higher levels of performance. Many hand-tuned, parallel
libraries dedicated to LQCD, such as Chroma \cite{chroma} or QUDA \cite{clark}
propose building blocks for these algorithms, so as to ease their
development. Parallelism does not stem from the algorithm themselves
but from the structure of the sparse matrices involved in the
computation. However, there are a number of shortcomings to the
library approach: First, any significant evolution of the hardware
requires to tune the library to the new architecture. This can lead to
change the grain of parallelism (as an adaptation to cache sizes for
instance) or change the data layout. Then, sparse matrices for LQCD,
in particular the matrix $A$, have a very regular structure
and library functions take advantage of it. Combining different
algorithms, such as preconditioners, leads to structural changes
in matrices that are not supported by these libraries. These two
limitations hinder considerably the time necessary to develop a
parallel/distributed code.

We propose in this paper a high-level Domain-Specific Language, QIRAL,
to model LQCD problems and enable automatic parallel code
generation. This novel approach makes possible the exploration and
test of new algorithms for LQCD. Sparse matrices in LQCD can be structured using algebra operators on dense matrices and the key idea is to use this structure to express parallelism. 
This extends ideas proposed in SPIRAL \cite{spiral},
the library generator for DSP algorithms to more complex codes and
matrices.
The problem formulation is first presented in Section \ref{sec:lqcd},
then the language is described in Section \ref{sec:qiral} and its
implementation in Section \ref{sec:compile}.

\section{Lattice QCD Description}
\label{sec:lqcd}
The principle of LQCD is to discretize space-time and describe the
theory on the resulting lattice. The lattice has to contain a very
large number of sites since it has to be fine grained and describe
large enough volumes. LQCD exists since 1974. Huge progress in
hardware, software and algorithms have been achieved, but it is not
enough to really sit on the parameters of nature. The light quarks,
\dquote{u} and \dquote{d} are still described by quarks heavier than in nature while
the heavy \dquote{b} quark is described as lighter. This implies systematic
errors in our results. To break this limitation, several orders of
magnitudes will be needed in the computing power and related
resources. It will demand new hardware with several level of
parallelism, and make coding more and more complicated.  Our goal is
to provide tools helping to face this new situation.

The heaviest part in the LQCD calculation is to generate a large
Monte-Carlo sample of very large files (field configurations)
according to an algorithm named \dquote{Hybrid Monte-Carlo}. It is a
Markovian process, every step of which takes several hours on the most
powerful computers. The algorithm is complex but it spends most of its
time in inverting very large linear systems which depend on the field
configuration.  Typically one will deal with matrices with billions of
lines and columns. The second heaviest part in the calculation is to
compute "quark propagators" which again boils down to {\it solving
  large linear systems} of the same type\cite{luscher01}. Therefore we
will {\it concentre on this task.}  The matrices involved in these
computation represent 4D stencil computations, where each vertex is
updated by the value of its 8 neighbours. Their structure is therefore
very regular, statically known, and it is used in the following
section.

\section{A High-Level Domain-Specific Language}
\label{sec:qiral}
QIRAL is a high level language for the description of algorithms and
for the definition of matrices, vectors and equations specific to
LQCD. The objective of the algorithmic part is to define algorithms
independently of the expression of sparse matrices used in LQCD. The
objective of the system of definitions and equations is to define
properties and structure sparse matrices in order to be able to find
parallelism.  While the algorithmic part uses straightforward
operational semantics, the equational part defines a rewriting system.

For the sake of simplicity, QIRAL is a subset of \LaTeX{}, meaning that the
QIRAL input can be either compiled into a \textit{pdf} file, for
rendering purposes, or compiled into executable codes. Algorithms and
definitions correspond to different predefined \LaTeX{} environments.

\paragraph{Variables, Types:}
Variables and constants used in QIRAL are vectors (denoted by the type
\texttt{V}) of any length, matrices of any size (\texttt{M}), complex
($\Complex$) or real numbers ($\Real$).  Besides, counted loops are
indexed by variables of type \texttt{index}, iterating over a domain
(denoted \texttt{indexset}). Index variables are potentially
multidimensional, and integers are a particular case of index value.
Functions of any number of argument of these types can be also
defined.
The size of a vector is defined through the size of its index set. The
size of an index set can be left undefined. If \texttt{V1} is the
vector and \texttt{IS} is an index set, \texttt{V1[IS]} denotes the
subvector indexed only by \texttt{IS}.

Matrices and vectors cannot be manipulated or defined
element-wise. Instead, matrices and vectors are built using either
constant, predefined values such as identity $I_{IS}$ (for the
identity on an index set $IS$), or operators such as $+,-,*$ and the
transposition, conjugate, direct sum and tensor product. The tensor
product $\otimes$ and direct sum $\oplus$ are defined by:
{\tiny
\[A \otimes B = [a_{ij}B]_{ij},\qquad A \oplus B = \left(\begin{array}{cc}A&0\\ 0&B\end{array}\right).
\]}
These two operators define parallel operations and help defining the
sparse matrices of LQCD.

\paragraph{Definition and Equations:}
All LQCD knowledge is given as a system of definition (for matrices
and vectors) and a system of equations on these values. In particular,
the structure of sparse matrices, explicitly defining where non-zero
elements are located, is given in these definitions. This structure is
then propagated in the algorithms.  Figure~\ref{eq:def} defines the
sparse Dirac matrix. Declarations of index sets $L$, $C$, $S$, $D$ are
not shown here, as well as the declaration of the constant matrices
used here.

\begin{figure}
{\tiny
\begin{align*}
  \Dirac
  & = \id{L \tensor C \tensor S} \\
  & + 2 * i * \kappa * \mu * \id{L \tensor C} \tensor \gamma_5 \\
  & + \kappa * \ssum{d \in  D}{((\shift{L}{- d} \tensor \id{C}) * \dsum{s \in L}{\U(d)[s]}) \tensor (\id{S} + \gamma[d])} \\
  & + \kappa * \ssum{d \in  D}{((\shift{L}{d} \tensor \id{C}) * \dsum{s\in L}{\U(- d)[s]}) \tensor (\id{S} - \gamma[d])} \;
 \end{align*}
}
\caption{\label{eq:def}Definition of Dirac matrix  on a Lattice $L$ in QIRAL.}
\end{figure}

The direct sums indexed by the vertices of the lattice ($L$) define
block-diagonal matrices. These diagonals are then shifted by $d$ or
$-d$ columns by a permutation matrix $\shift{L}{d}$ (predefined).

\paragraph{Algorithms, Preconditioners:}
Many algorithms proposed for solving Equation \ref{eq:dirac} are
Krylov methods. Figure~\ref{fig:algo} presents two algorithms in
QIRAL. The first is a variant of the conjugate gradient and is representative of iterative methods. The second one, Schur complement method, is a preconditioner: Computing $x$, solution of Equation~\ref{eq:dirac} is achieved by computing two solutions to smaller problems. 

The keyword \texttt{Match} helps to define the algorithm as a
rewriting rule: Whenever the statement in the \texttt{Match} condition
is found, it can be rewritten by the algorithm. For the \texttt{SCHUR}
preconditioner, this rewriting can be performed at the
condition that the requirement is fulfilled: $P_1 A P_1^t$ has to be
invertible. Using identities defined through the equation system, the
rewriting system proves here that indeed, the requirement is valid,
when matrix $A$ is Dirac matrix and $P_1$ is a projection matrix
keeping only vertices of the lattice with even coordinates (even-odd
preconditioning). 
\begin{figure}[h!]
\begin{minipage}[t]{0.48\textwidth}
\tiny
\begin{template}{CGNR}
\Input{$A\in M , b \in V,\epsilon \in \Real$}
\Output{$x \in V $}
\Match{$x = A^{-1} * b \;$}
\Var{$r, p, Ap, z \in V , \alpha, \beta, n_r, n_z, n_{z1}\in \Real$}
     	$r = b$ \;
	$z = A ^\dagger * r$\; 
	$p = z$ \; 
        $x = 0$ \;
	$n_z = (z \hdp z)$ \;
        $n_r = (r \hdp r)$ \;
	\While{$(n_r > \epsilon)$} {
	  $Ap = A * p $\;
	  $\alpha = n_z / (Ap \hdp Ap)$ \;
	  $x  = x + \alpha * p$ \;
	  $r = r - \alpha * Ap$ \;
	  $z = A ^\dagger * r$ \;
	  $n_{z1} = (z \hdp z)$ \;
	  $\beta = n_{z1} / (n_z)$ \;
	  $p = z + \beta * p$ \;
	  $n_z = n_{z1}$ \;
          $n_r = (r \hdp r)$ \;
	}
\caption{Conjugate Gradient, Normal Resolution [CGNR]}
\end{template}
\end{minipage}
\begin{minipage}[t]{0.52\textwidth}
\tiny
\begin{template}{schur}
\caption{Definition of Schur complement method [SCHUR]}
\Input{$A, P_1, P_2 \in M , b \in V $} \Output{$x \in V $}
\Match{$x = A^{-1} * b \;$} \Var{$v_1, v_2, x_1, x_2 \in V, D_{11},
  D_{12}, D_{21}, D_{22} \in M$}
\Require{$\isInvertible{P_1 * A * P_1 ^t}$}
$D_{21} = P_2 * A * P_1^t$ \; 
$D_{11} = P_1 * A * P_1^t$ \; 
$D_{22} = P_2 * A * P_2^t$ \; 
$D_{12} = P_1 * A * P_2^t$ \; 
$v_1 = P_1 * b$ \;
$v_2 = P_2 * b$ \; 
$x_2 = (D_{22} - D_{21} * D_{11}^{-1} * D_{12})^{-1} * (v_2 - D_{21} * D_{11}^{-1} * v_1)$ \; 
$x_1 = D_{11}^{-1} * (v_1 - D_{12} * x_2) $ \; $x = P_1^t * x_1 + P_2^t * x_2$ \;
\end{template}
\end{minipage}
\caption{Two algorithms. On the left: A variant of the conjugate gradient method; on the right: A preconditioner, the Schur complement method.}\label{fig:algo}
\end{figure}
QIRAL is expressive enough to represent different variants of
conjugate gradient, BiCGSTAB, methods with restart (such as GCR), and
other methods proposed by physicists.

\section{From QIRAL to parallel code}
\label{sec:compile}
\begin{figure}[h]
\centering
\includegraphics[width=0.7\textwidth]{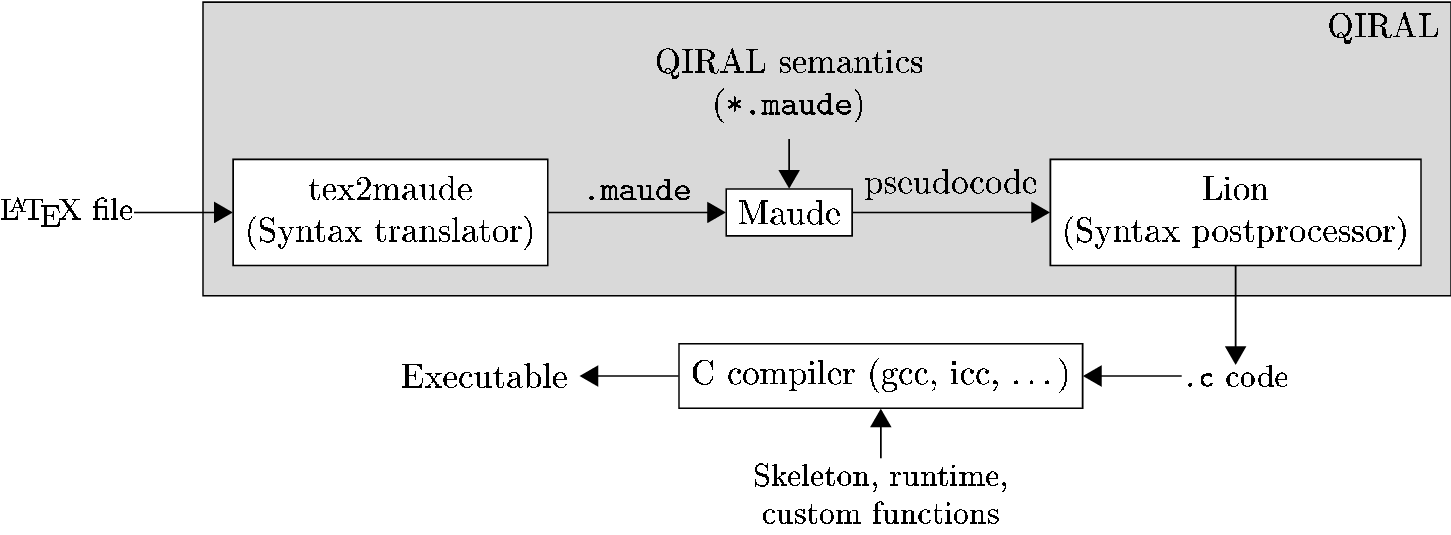}
\caption{Overview of QIRAL compilation chain}\label{fig:overview}
\end{figure}

The QIRAL compiler takes as input a file describing algorithms and
equations and generates parallel code. The phases are described in
\ref{fig:overview} and presented in detail in the following.

\paragraph{Rewriting System Generation:}
From its current \LaTeX{} implementation, the QIRAL input is
translated into a set of rewriting rules and equations, using Maude
framework~\cite{maude}. Maude is a multi-purpose rewriting system,
handling equational rewriting rules and reflection. More precisely,
definitions are translated by a first phase, \texttt{tex2maude} into
equations (or conditional equations) and algorithms into rewriting
rules. The output file contains domain specific definitions (e.g. the
Dirac operator from Figure~\ref{eq:def}), the desired program (in
our case $x=Dirac^{-1} b$), and all algorithms and preconditioners we
want to apply.

Definitions have a semantic similar to a rewriting rule: the left hand
side is rewritten into the right hand side. To ensure the existence of
a normal form, the system of equation has to be convergent and
confluent. So far, our implementation does not use a tool such as an
automatic Church-Rosser Checker \cite{crc} .  Algorithms are
translated into conditional rules, applied is their prerequisite are
checked. The left hand side of the rule correspond to the
\texttt{Match} clause and the algorithm itself is the right hand side.

This rewriting system is merged with another one defining general
algebraic properties, code generation rewriting rules and code
optimizations. The main phase of the QIRAL compiler is therefore
described through rewriting system, following works such as
Stratego~\cite{stratego}. QIRAL is a static strongly type language,
vectors and matrices are defined by their index sets. Type checking is
the first analysis achieved by this main phase.

\paragraph{Applying Algorithms and Simplifications:}
The user provides the list of algorithms to compose and the initial
program (here, we focused on the equation $x = Dirac^{-1}
b$). Exploring different algorithms and sequences of algorithms boils
down to change the list given to the QIRAL compiler. Checking
that algorithm requirements is automatically achieved
by the rewriting system, using the equational theory provided through
the system of definitions. The equational system, using both equations coming from LQCD definitions and algebraic properties involving the different operators, simplifies terms that are equal to zero. 

The result is a program where the matrix $A$ of the algorithms has been
replaced either by the Dirac matrix, or by a matrix obtained through
transformation by preconditioners.

\paragraph{Loop generation and parallelization:}
At this step, the initial statement has been replaced by the algorithm
statements, directly using the Dirac matrix or a preconditioned
version. Assignment statements are vector assignments: Values for all
vertices of the lattice can be modified in one statement. Matrices are
still described using tensor products and direct sums. 

Loops are obtained by transforming all indexed sums, products into
loops or sequences. For instance, the direct sum operator indexed by
the lattice $L$ in the definition of Dirac matrix, in Figure
\ref{eq:def} is transformed into a parallel loop over all elements of
the lattice. This loop is parallel, due to the meaning of $\oplus$. As
the lattice is 4D, either 4 nested loops are created, or one single,
linearized loop is created. The choice depends on a parameter in the
QIRAL compiler.

Some usual compiler transformations are then used, computing
dependences, fusioning loops, applying scalar promotion and other
optimizations. At the end of this phase, an OpenMP code is produced,
essentially by identifying for parallel loops the set of private
variables. These transformations are driven by rewriting strategies,
using reflection in Maude. 

\label{sec:details}

\paragraph{Matching Library Calls:} The resulting code still uses some high level operators on dense matrices, such as tensor products on dense matrices, matrix-vector product. These operators are then replaced by library calls. For this step, it is sufficient to define for each library the expression it computes, as a rewriting rule. We defined LION, a set of hand-written library functions used for validation purposes. 


\begin{figure}[h]
\centering
{\tiny
\begin{verbatim}
...
while(nr  > epsilon) {      
 #pragma omp parallel for private(ID30,ID31,ID43,ID44,ID56) 
 for(iL = 0; iL < L; iL ++) {
  ID31 = spnaddspn(matmulspn(tensor(U[uup(iL, dt)], gmsubgm(gmdiag(c1), G(dt))), p[sup(IDX, dt)]), 
   spnaddspn(matmulspn(tensor(U[uup(iL, dz)], gmsubgm(gmdiag(c1), G(dz))), p[sup(IDX, dz)]), 
   spnaddspn(matmulspn(tensor(U[uup(iL, dy)], gmsubgm(gmdiag(c1), G(dy))), p[sup(IDX, dy)]), 
   matmulspn(tensor(U[uup(iL, dx)], gmsubgm(gmdiag(c1), G(dx))), 
   p[sup(IDX, dx)])))) ; 
  ID30 = cplmulspn(kappa, ID31) ; 
  ID44 = spnaddspn(matmulspn(tensor(U[udn(iL, dt)], gmaddgm(gmdiag(c1), G(dt))), p[sdn(IDX, dt)]), 
   spnaddspn(matmulspn(tensor(U[udn(iL, dz)], gmaddgm(gmdiag(c1), G(dz))), p[sdn(IDX, dz)]), 
   spnaddspn(matmulspn(tensor(U[udn(iL, dy)], gmaddgm(gmdiag(c1), G(dy))), p[sdn(IDX, dy)]), 
   matmulspn(tensor(U[udn(iL, dx)], gmaddgm(gmdiag(c1), G(dx))), p[sdn(IDX, dx)])))) ; 
  ID43 = cplmulspn(kappa, ID44) ; 
   ...
\end{verbatim}
}
\caption{Sample output C code produced for CGNR
  algorithm}\label{fig:pseudo_code}
\end{figure}

\begin{figure}[ht]
\centering
\includegraphics[angle=-90,width=0.5\textwidth]{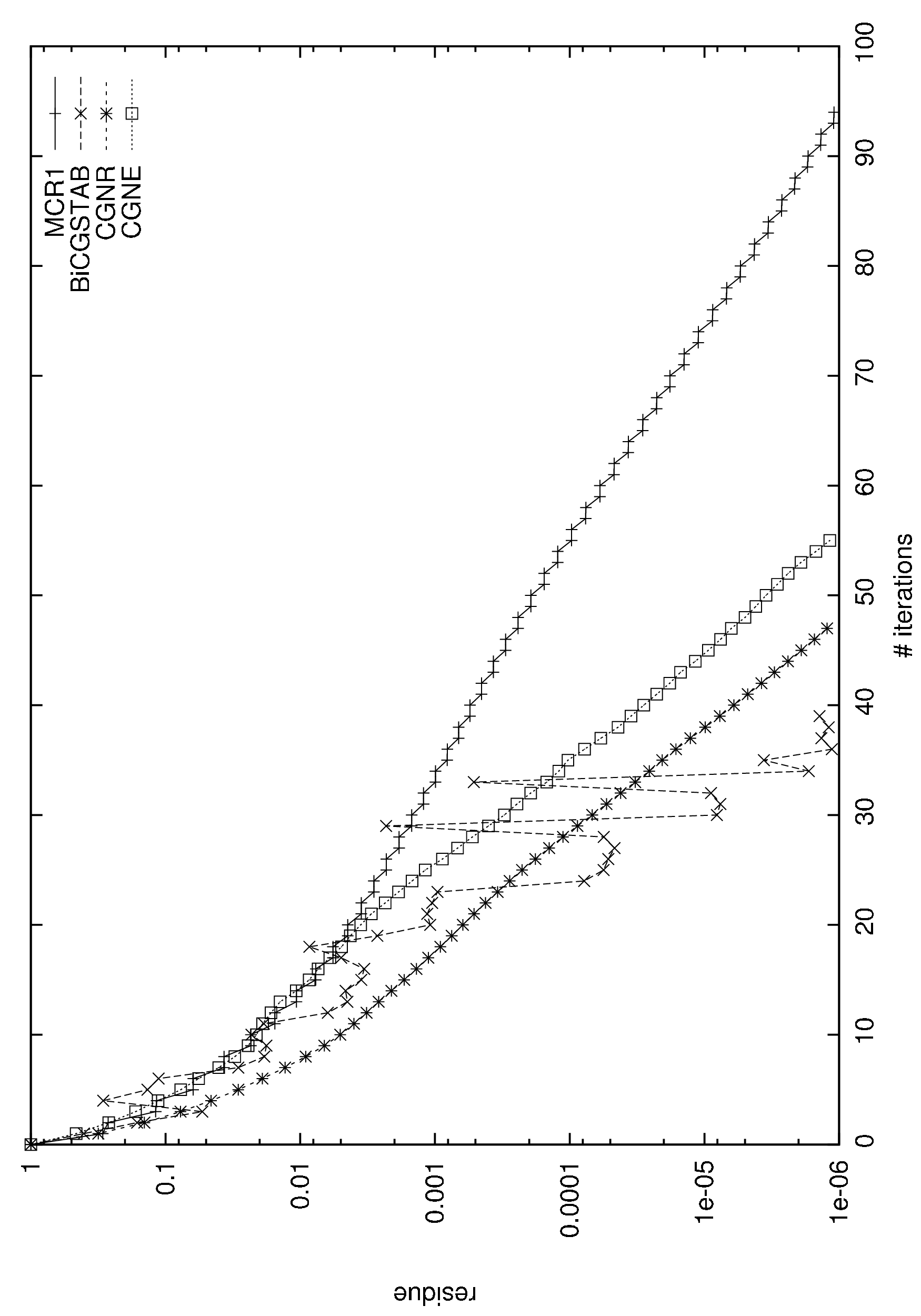}
\caption{\label{fig:iteration}Convergence speed for a lattice of size $4^4$ for different methods: Conjugate Gradient Normal Error (CGNE), Normal Resolution (CGNR), Biconjugate Gradient (BiCGSTAB), Modified conjugate gradient with preconditioning (MCR1). }\end{figure}

Finally, post-processing phase, mostly syntactic rewriting, transforms
this output into a C function. This function is compiled with a
program skeleton and a runtime that initializes data, calls the
generated code and stores the result. Custom functions can also be
defined here since undefined operations from the QIRAL input appear as
functions calls. One application is to call BLAS routines instead of
letting QIRAL implement them. For instance one defines a rule that
says $(C = A * B) = \text{\texttt{dgemm}}(A, B, C)$.
At this stage, we are more concerned about the correctness of the
output rather than the efficiency of the code, which will be the
purpose of further steps. Automatically generated codes for different
algorithms and preconditionings show different convergence speed, as
shown in Figure \ref{fig:iteration}.

\section{Conclusion}
\label{sec:conclusion}
This paper has presented an overview of QIRAL, a high level language
for automatic parallel code generation of Lattice QCD codes. The
language is based on algorithmic specifications and on mathematical
definitions of mathematical objects used in the
computation. 

The contribution of this short paper is to show that a high-level
representation makes possible the automatic generation of complex LQCD
parallel code. Parallelism directly stems from the structure of
sparse, regular matrices used in LQCD. While the initial matrix
represents a stencil computation, QIRAL is able to manipulate more
complex structures obtained through preconditioning for instance,
unlike Pochoir~\cite{pochoir}. The approach, similar to the one
proposed by Ashby \textit{et al.}~\cite{ashby} enables the user to
define new equations and domain-specific definitions. The QIRAL
compiler is able to keep such information through 
transformations resulting from preconditioners or algorithms. The tensor
product and direct sum operators are translated into parallel
loops and lead to OpenMP code generation.  This way, algorithmic
exploration, the key for higher levels of performance, can be freed
from the constraints and costs of parallel tuning. Besides generation
of distributed codes with communications is within reach. For
heterogeneous architectures, such as Cell or GPUs, further work for
automatic data-layout optimization is required, in order to reach
levels of performance of previous works (such as
\cite{vranas,khaled01,tad2010} for the CELL BE and \cite{clark} for
the GPU).


\bibliographystyle{plain}
\bibliography{paper}

\begin{thebibliography}{10}

\bibitem{ashby}
T.~Ashby, A.~Kennedy, and M.~O'Boyle.
\newblock {Cross Component Optimisation in a High Level Category-Based
  Language}.
\newblock In Marco Danelutto, Marco Vanneschi, and Domenico Laforenza, editors,
  {\em Euro-Par Parallel Processing}, volume 3149 of {\em LNCS}, pages
  654--661. Springer, 2004.

\bibitem{clark}
M.A. Clark, R.~Babich, K.~Barros, R.C. Brower, and C.~Rebbi.
\newblock {Solving lattice QCD systems of equations using mixed precision
  solvers on GPUs}.
\newblock {\em Computer Physics Communications}, 181(9):1517 -- 1528, 2010.

\bibitem{maude}
Manuel Clavel, Fransisco Dur\'{a}n, Steven Eker, Patrick Lincoln, Narciso
  Mart\'{\i}-Oliet, Jos\'{e} Meseguer, and Jose~F. Quesada.
\newblock {The Maude System}.
\newblock In {\em Intl. Conf. on Rewriting Techniques and Applications}, pages
  240--243, London, UK, 1999. Springer-Verlag.

\bibitem{crc}
Francisco Dur\'an and Jos\'e Meseguer.
\newblock A church-rosser checker tool for conditional order-sorted equational
  maude specifications.
\newblock In {\em Rewriting Logic and Its Applications}, volume 6381 of {\em
  LNCS}, pages 69--85. Springer, 2010.

\bibitem{chroma}
Robert~G. Edwards and Balint Joo.
\newblock {The Chroma Software System for Lattice QCD}.
\newblock {\em NUCL.PHYS.PROC.}, 140:832, 2005.

\bibitem{khaled01}
Khaled~Z. Ibrahim and Francois Bodin.
\newblock {Implementing Wilson-Dirac operator on the cell broadband engine}.
\newblock In {\em Intl. Conf. on Supercomputing}, pages 4--14, New York, NY,
  USA, 2008. ACM.

\bibitem{luscher01}
Martin L\"uscher.
\newblock Local coherence and deflation of the low quark modes in lattice qcd.
\newblock {\em J. of High Energy Physics}, 2007(07):081, 2007.

\bibitem{spiral}
Markus P{\"u}schel, Franz Franchetti, and Yevgen Voronenko.
\newblock {\em Encyclopedia of Parallel Computing}, chapter Spiral.
\newblock Springer, 2011.

\bibitem{tad2010}
Claude Tadonki, Gilbert Grodidier, and Olivier Pene.
\newblock {An efficient CELL library for lattice quantum chromodynamics}.
\newblock {\em SIGARCH Comput. Archit. News}, 38:60--65, January 2011.

\bibitem{pochoir}
Yuan Tang, Rezaul~Alam Chowdhury, Bradley~C Kuszmaul, Chi-Keung Luk, and
  Charles~E Leiserson.
\newblock {The pochoir stencil compiler}.
\newblock {\em Symp. on Parallelism in algorithms and architectures}, page 117,
  2011.

\bibitem{stratego}
Eelco Visser and Zine el~Abidine~Benaissa.
\newblock A core language for rewriting.
\newblock {\em Electronic Notes in Theoretical Computer Science}, 15(0):422 --
  441, 1998.
\newblock Rewriting Logic and its Applications.

\bibitem{vranas}
Pavlos Vranas, Matthias~A. Blumrich, Dong Chen, Alan Gara, Mark Giampapa,
  Philip Heidelberger, Valentina Salapura, James~C. Sexton, Ron Soltz, and Gyan
  Bhanot.
\newblock Massively parallel quantum chromodynamics.
\newblock {\em IBM J. of Research and Development}, pages 189--198, 2008.

\bibitem{qcd}
Frank Wilczek.
\newblock { What QCD Tells Us About Nature -- and Why We Should Listen}.
\newblock {\em NUCL.PHYS.A}, 663:3, 2000.

\end{thebibliography}
\end{document}